\documentclass[a4paper,11pt]{article}
\usepackage{pos}
\usepackage{caption}
\usepackage{subcaption}
\usepackage{aas_macros}
\usepackage{gensymb,multirow,wrapfig,lineno,enumitem}

\title{Cosmic inventory of the background fields of relativistic particles in the Universe}
\ShortTitle{Cosmic inventory of the background fields of relativistic particles in the Universe}

\author*[a,b]{Jonathan~Biteau}
\affiliation[a]{Universit\'e Paris-Saclay, CNRS/IN2P3, IJCLab, Orsay, France}
\emailAdd{biteau@in2p3.fr}
\affiliation[b]{Institut universitaire de France (IUF), France}

\abstract{The extragalactic background is composed of the emission from all astrophysical sources, both resolved and unresolved, since the epoch of reionization, in addition to any diffuse components of exotic nature. In the last decade, there has been significant progress in our understanding of the cosmic history of extragalactic emissions associated with stellar evolution and accretion onto supermassive black holes, largely enabled by the extensive body of multi-wavelength data. The brightness of the extragalactic sky is now measured in photons, neutrinos, and cosmic rays, using observatories on the ground, in the sea, and in the ice, satellites in Earth orbit, and probes at the edge of the solar system. This wealth of disparate data is essential to unraveling the mysteries of the source populations that contribute to the extragalactic background and understanding the origin of their power.

In this contribution, we present an open database containing the most comprehensive collection of multi-messenger measurements of the extragalactic background spectrum to date. The convergence of direct measurements, galaxy counts, and indirect measurements is remarkable, except in the radio band where an observational controversy remains. The combination of multi-messenger measurements over 27 frequency decades allows us to estimate the energy density of most extragalactic background components with an uncertainty of less than 30\%. We explore the consistency of this cosmic inventory of the observed fields of relativistic particles populating the Universe with the cosmic history of star formation and accretion around supermassive black holes. Models incorporating these cosmic histories, as well as the redshift-dependent luminosity functions of extragalactic sources, currently match the electromagnetic component of the extragalactic background spectrum over 14 frequency decades, from the near UV to sub-TeV gamma rays. The knowledge gained from synthetic population models in the electromagnetic bands may become a crucial tool for understanding the origin of the most energetic extragalactic messengers, neutrinos and ultrahigh-energy cosmic rays.}

\FullConference{39th International Cosmic Ray Conference (ICRC 2025)\\
 15–24 July 2025 \\
Geneva, Switzerland\\}

\begin{document}
\maketitle

\section{Introduction}

The surface brightness of the night sky is a fundamental observable in astrophysics and cosmology. After subtracting local foregrounds, such as the Milky Way or sunlight reflected by interplanetary dust, measuring the brightness of the extragalactic sky yields the radiative balance of all emission processes in the Universe \cite{2021arXiv210212089D}. The observed specific intensity, $I_\nu$ in nW\,m$^{-2}$\,sr$^{-1}$\,Hz$^{-1}$ or $E\, J(E)$ in m$^{-2}$\,s$^{-1}$\,sr$^{-1}$, is a direct tracer of the energy density of isotropic, relativistic particle fields in cosmic voids, $u_\nu = \frac{4\pi}{c}I_\nu$ in eV\,m$^{-3}$\,Hz$^{-1}$. In other words, extragalactic sky brightness provides the integral energy density of photons, neutrinos, and cosmic rays emitted at all cosmic epochs. This integral of cosmic emissivity up to zero redshift traces the history of all emission processes of astrophysical origin, including nucleosynthesis and emission by dust associated with star formation, as well as the accretion and ejection of plasma around supermassive black holes. This extragalactic background observed across wavelengths and messengers also offers a probe of fundamental physics processes, through the decay or annihilation of dark matter candidates.

The brightness of the night sky is a pillar of modern cosmology \cite{1993QJRAS..34..157L}. The universally observed fact that the night sky is less bright than the surface of an average star, such as the Sun, conflicts with the idea of a stationary universe with an infinite number of sources. The resolution of this paradox — the finite history of source formation — was not widely accepted by the scientific community until the 1990s, although Lord Kelvin had validated this hypothesis at the dawn of the 20th century \cite{1986Natur.322..417H}. Our current understanding of the extragalactic background is the result of a long scientific history whose origins can be traced back to Digges in 1576, following the Copernican Revolution. This history has been marked by the tribulations of astronomers such as Halley, de Ch\'eseaux, Olbers, and Herschell, as well as great names in mathematics, philosophy, and literature (see \cite{1990IAUS..139....3H} for an historical overview).

This fundamental field of study, once considered metaphysical, has seen tremendous observational advances since the end of the 20th century. We have moved from a regime largely dominated by upper limits due to the difficulty of subtracting foregrounds to a vast volume of data from ground-based observatories, satellites, and interstellar probes. These measurements cover wavelengths ranging from tens of meters (20\,MHz) to attometers ($10^{12}\,\rm{eV} \equiv 1$\,TeV) and now extend to the $\rm{PeV}-\rm{ZeV}$ range ($10^{15}-10^{21}\,$eV) for neutrinos and atomic nuclei that constitute ultra-high-energy cosmic rays. In line with the seminal work of \cite{1972SvPhU..14..569L, 1990ComAp..14..323R}, the purpose of these proceedings is to review the state of the art in measurements, limits, and possible tensions on extragalactic background radiation. Here, we mainly focus on the value of the monopole, which is the sum of the energy fluxes per unit solid angle from resolved and unresolved sources as a function of frequency, $\nu$. The monopole of the universe spectrum is represented below in the form of $\nu I_\nu$ as a function of $\log_{10}(\nu)$. From this representation, we can derive the bolometric intensity, $I = \int \mathrm{d}\nu\, I_\nu = \ln(10) \int \mathrm{d}\log_{10}(\nu)\, \nu I_\nu$, and the energy density,  $u = \frac{4\pi}{c} I$, of the radiation fields.

\section{Datasets: from radio wavelengths to ultra-high-energy cosmic rays}

The extragalactic background synthesis presented here builds on the data collection of \cite{2018ApSpe..72..663H}. We supplemented these data with a bibliographic review, up to the end of 2024, aimed for exhaustive, non-redundant coverage. The 70 spectra from the literature (including 33 spectra from \cite{2018ApSpe..72..663H}) are stored in Enhanced Character-Separated Values (ECSV) format, supported by Astropy \cite{2022ApJ...935..167A}. These data and the associated software are publicly available on GitLab and Zenodo \cite{biteau_2023_7842239}.
%\footnote{\url{https://gitlab.com/jbiteau.pro/the_multimessenger_extragalactic_spectrum}}. 
The display software is loosely based on \cite{evoli_carmelo_2020_4396125}, while the analysis software is custom-made. Each ECSV file contains four columns: frequency or energy, intensity or temperature, and the lower and upper uncertainties on the intensity or temperature (at the 68\% confidence level, including systematic uncertainties when provided by the authors). The metadata describing the columns and their units allow for a relatively straightforward conversion from one unit system to another using the Astropy equivalence scheme (see also \cite{2018ApSpe..72..663H} for unit conventions). The metadata contains information about the origin of the spectra, including the bibliographic reference in NASA/ADS format, the observatory from which the spectrum originated, and the type of observable. Three broad categories of observables have been chosen: integrated galaxy light (IGL) from galaxy counts, which can be considered lower limits insofar as they neglect low-surface-brightness emissions; integral measurements, which take into account both resolved sources and diffuse components; and direct observations in dark patches, most of which are considered upper limits due to possible contamination by foregrounds.

\begin{figure}[b]
    \centering
    \includegraphics[width=0.75\linewidth]{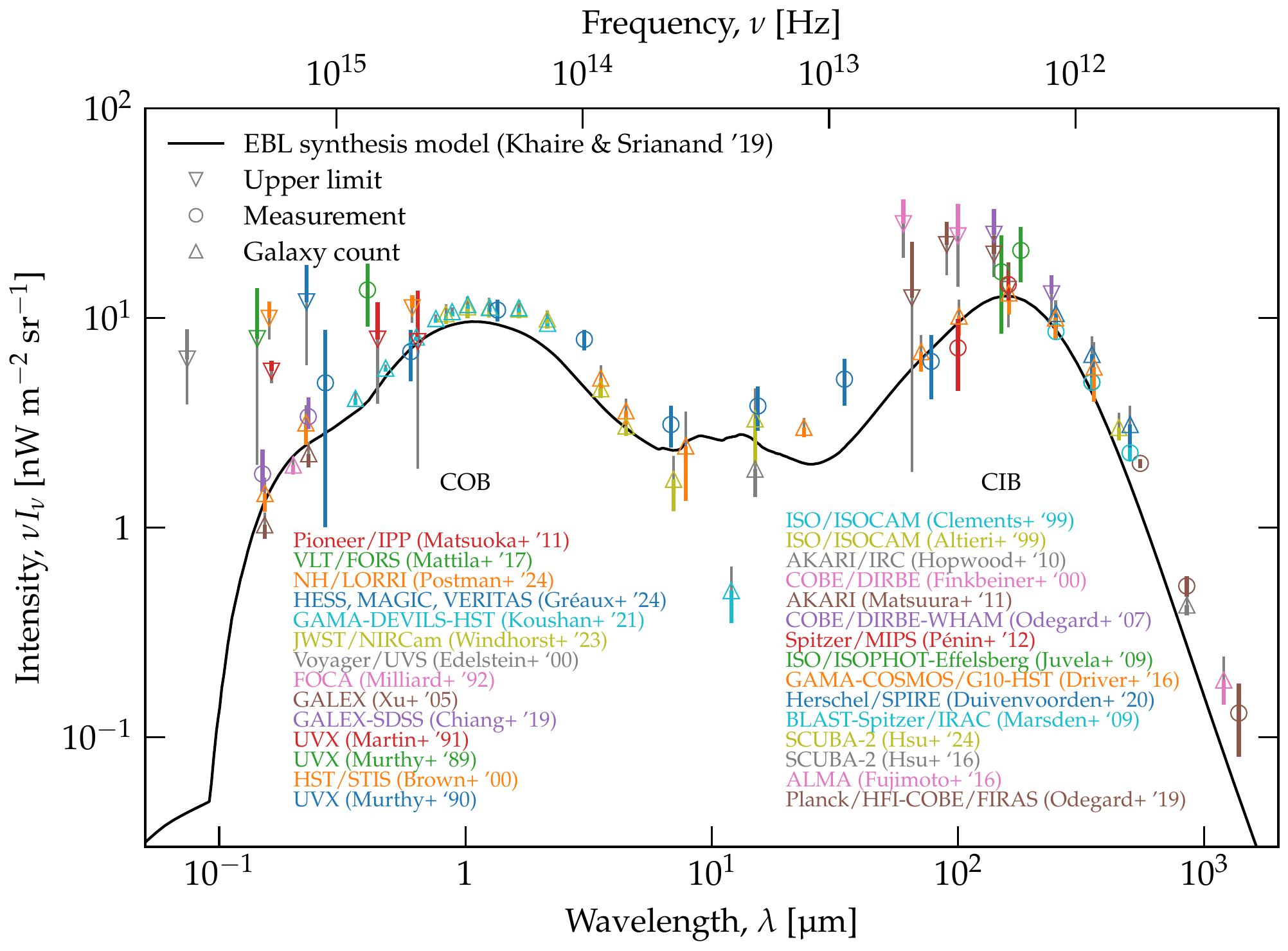}
    \caption{Spectrum of the Cosmic Optical and Infrared Backgrounds.}
    \label{fig:COIB}
\end{figure}

We discuss the components of the electromagnetic cosmic backgrounds below excluding the cosmic microwave background (CMB), which peaks at millimeter wavelengths with an amplitude of approximately 700\,nW\,m$^{-2}$\,sr$^{-1}$. The components are the cosmic optical and infrared backgrounds (COB, $0.1-8\,\mu$m; CIB, $8\mu\rm{m}\,-2\,$mm), which primarily originate from star-forming galaxies (SFGs); the cosmic X-ray and gamma-ray backgrounds (CXB, $0.3\,\rm{keV}-2\,$MeV; CGB, $2\,\rm{MeV}-1\,$TeV), mainly from active galactic nuclei (AGNs); and the cosmic radio background (CRB, $10\,\rm{MHz}-10\,$GHz), which has substantial contributions from both SFGs and AGNs. Consistent with the foundational work mentioned in the introduction, we also include the extragalactic neutrino background (ENB, $30\,\rm{TeV}-3\,\rm{PeV}$) and the extragalactic cosmic-ray background (ECRB, $0.2-200\,\rm{EeV}$), within their measured energy ranges.

\subsection{Cosmic Optical and Infrared Backgrounds}

Figure~\ref{fig:COIB} presents the spectral measurements and constraints obtained for the COB and CIB. Also shown is the population synthesis model of the extragalactic background light (EBL) from \cite{2019MNRAS.484.4174K}, intended to guide the reader. The IGL, including the extrapolation down to the faintest galactic fluxes, is determined with an accuracy of better than 20\% between 150\,nm and $450\,\mu$m, and of about 5\% between 350\,nm and $2.2\,{\mu}$m \cite{2021MNRAS.503.2033K}. The uncertainty in the IGL is primarily due to cosmic variance because of the limited number of fields covered by the observations. The integral measurements in Fig.~\ref{fig:COIB} include indirect estimates from gamma-ray absorption in the TeV range. These estimates are accurate to about 10\% in the micrometer range and to 30\% at 600\,nm and $80\,\mu$m \cite{2024ApJ...975L..18G}. Notable convergence is observed around 600\,nm between the IGL, indirect measurements, and dark patch observations made beyond Pluto's orbit \cite{2024ApJ...972...95P}. These dark-patch observations are presumably devoid of foreground contamination, and their estimated accuracy is 15\%. Convergence is also observed in the far infrared, beyond $200\,\mu$m, where contamination by Galactic cirrus and Zodiacal light is minimal. Outside these ranges, significant contamination by foregrounds is expected, which is likely responsible for the discrepancy between dark patch measurements and other datasets in Fig.~\ref{fig:COIB}.

\begin{wrapfigure}{R}{0.5\textwidth}
    \vspace{-0.4cm}
    \centering
    \includegraphics[width=\linewidth]{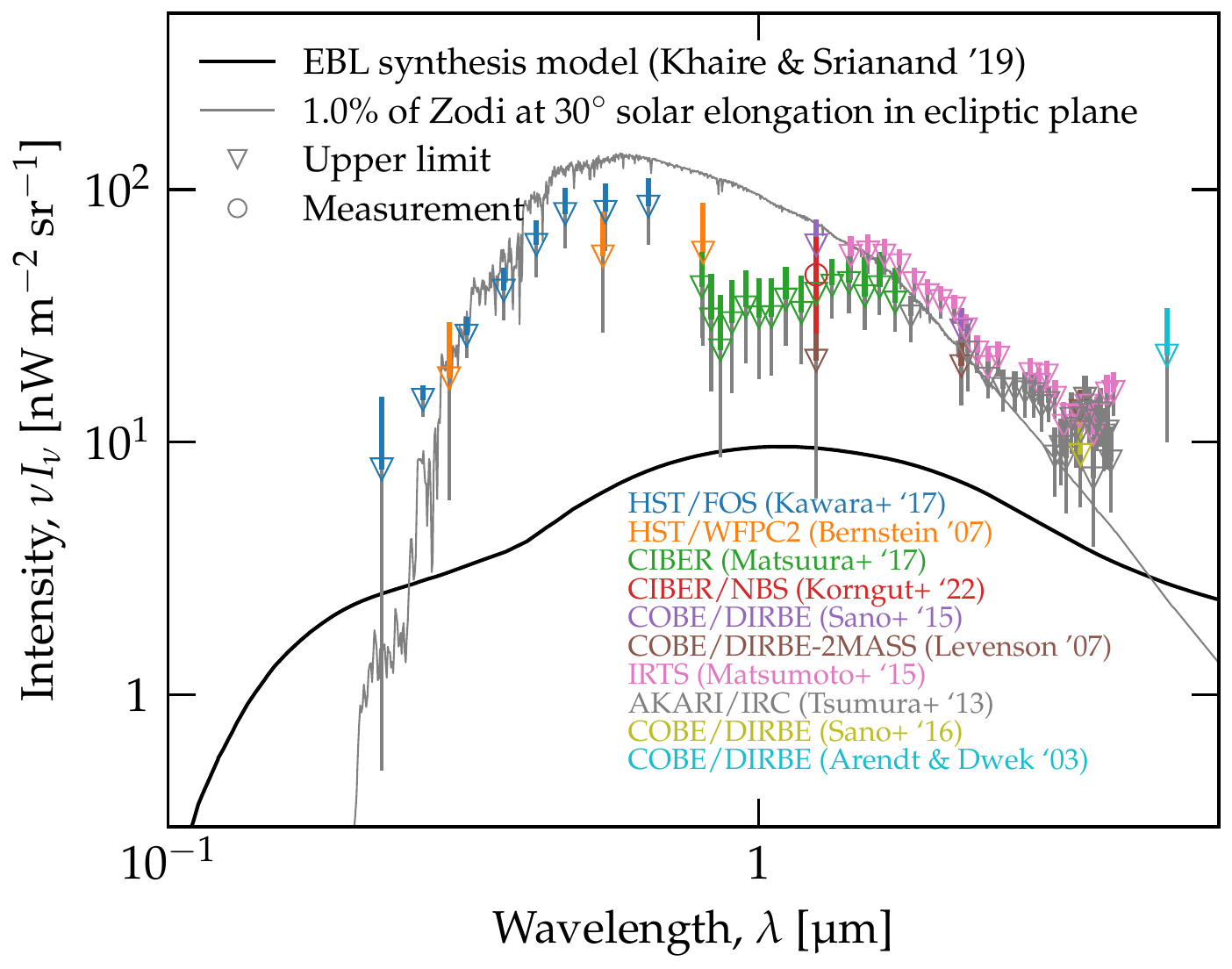}
    \caption{Dark-patch observations excluded from Fig.~\ref{fig:COIB}, possibly contaminated by Zodi.}
    \label{fig:Zodi}
    \vspace{-1cm}
\end{wrapfigure}
We deliberately excluded from the figure dark-patch observations that clearly differ from indirect and IGL measurements in the optical and near-infrared ranges. These observations are compared to the COB model of \cite{2019MNRAS.484.4174K} and to the solar spectrum normalized to 1\% of the Zodiacal light in Fig.~\ref{fig:Zodi}. Dark-patch observations can also be compared in this figure to the quasi-isotropic Zodi component (red circle) inferred by Fraunhofer spectroscopy around $1.25\,\mu$m \cite{2022ApJ...926..133K}. A significant confirmation of such an isotropic Zodi component, which is absent from the dominant zodiacal-light model by design, is still needed.

\begin{wrapfigure}{r}{0.5\textwidth}
    \vspace{-1.1cm}
    \centering
    \includegraphics[width=\linewidth]{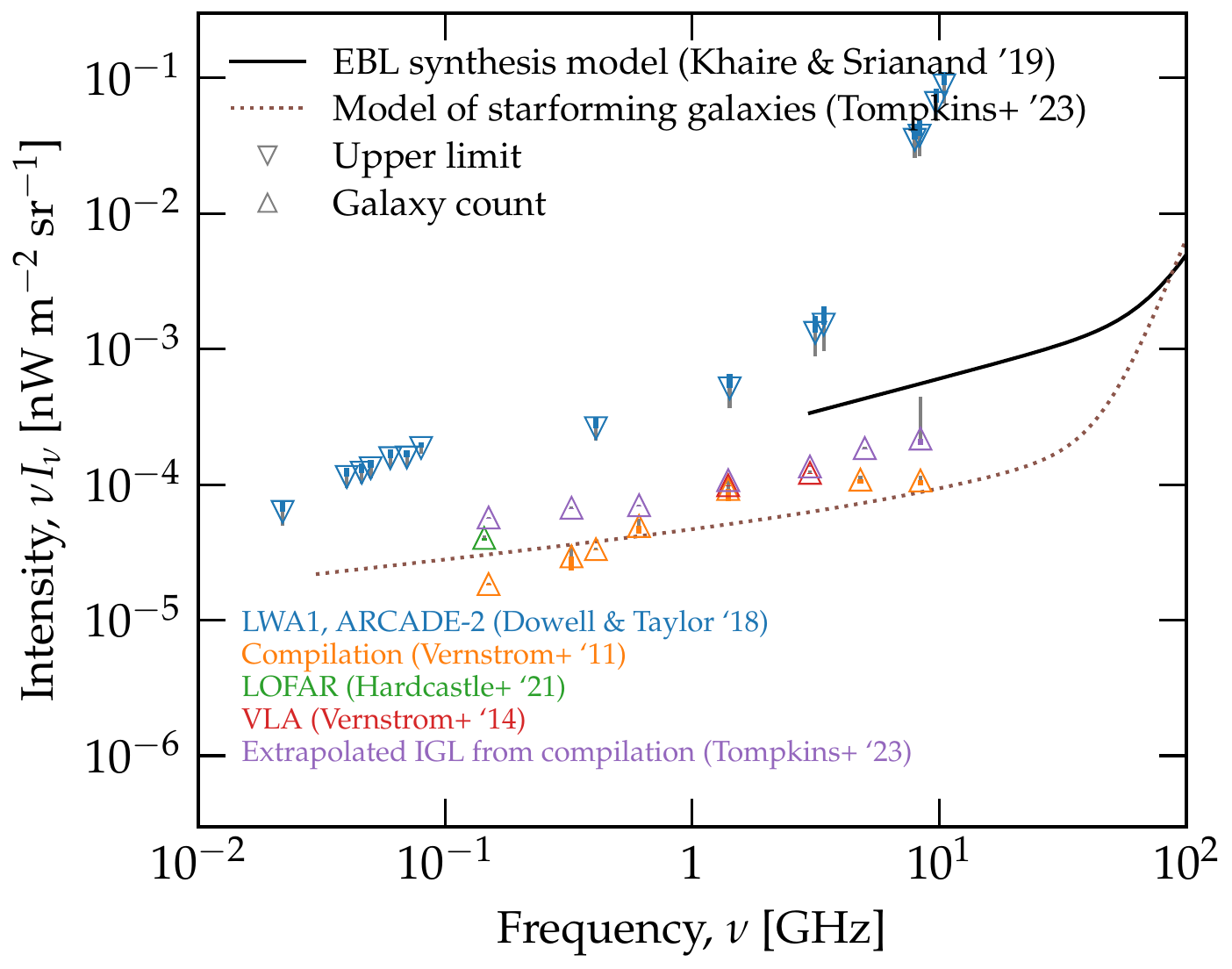}
    \caption{Spectrum of the Cosmic Radio Background.}
    \label{fig:CRB}
    \vspace{-1cm}
\end{wrapfigure}

\subsection{Cosmic Radio Background}

The extragalactic background is not well-constrained in the microwave range due to the dominance of the CMB. Below 10\,GHz, the CRB is constrained by IGL and dark-patch measurements, as shown in Fig.~\ref{fig:CRB}. The IGL converges in two frequency bands: 1.4\,GHz and 3\,GHz. These bands have roughly equivalent contributions from AGNs, luminous but low in number density, and SFGs, less luminous but more numerous. Dark-patch observations at high Galactic latitudes in these two bands are four and five times higher than the IGL, with tensions significant at less than $4\sigma$ and $3\sigma$, respectively. Whether this discrepancy is due to an additional CRB component or to foregrounds (e.g., the Galactic halo) remains an open question \cite{2018PASP..130c6001S}.

\subsection{Cosmic X-ray and Gamma-ray Backgrounds}

\begin{figure}[h]
    \centering
    \includegraphics[width=0.75\linewidth]{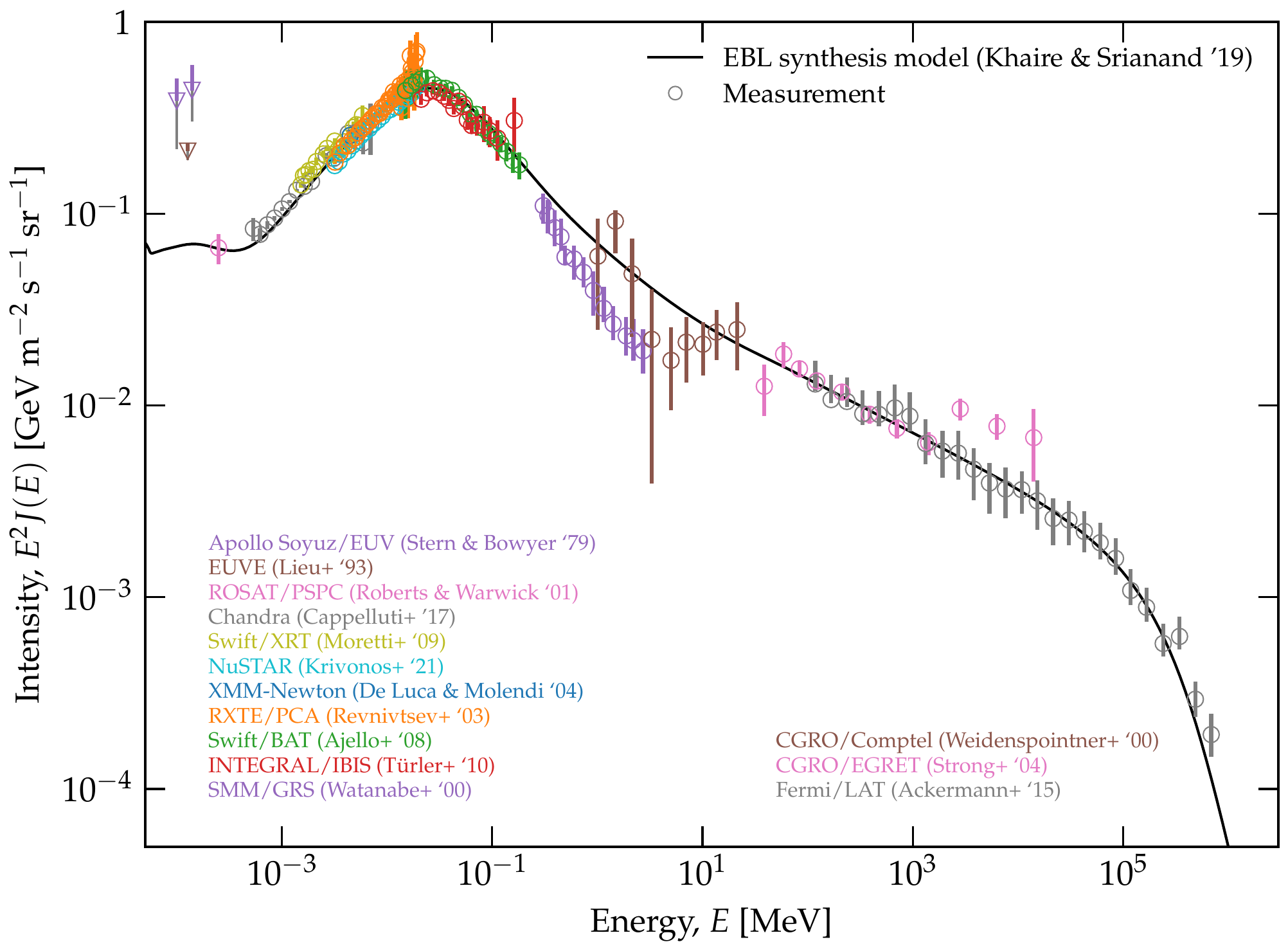}
    \caption{Spectrum of the Cosmic X-ray and Gamma-ray Backgrounds}
    \label{fig:CXGB}
    \vspace{-0.5cm}
\end{figure}

Along with the microwave range, the far UV to soft X-ray range is the other wavelength region of the EBL that is the least constrained observationally, due to absorption and emission by the Galactic foreground. At higher frequencies, the spectrum of CXB and CGB shown in Fig.~\ref{fig:CXGB} is well constrained.\footnote{The tensions between the spectra from CGRO and those from other observatories are below $<2\sigma$.} Measurements of the IGL and diffuse components largely converge in this range, attributing most of the CXB to AGNs without jets and most of the CGB to AGNs with jets. As discussed in \cite{2022hxga.book...78B}, deep surveys in the energy ranges of $0.2-2\,$keV, $2-8\,$keV, and $8-24\,$keV have resolved $80.9\% \pm 4.4\%$, $92.7\%\pm 13.3\%$, and nearly $35\%$ of the sources contributing to the CXB, respectively. At the highest photon energies, above 50\,GeV, nearly 70\% of the CGB has been resolved into point sources (including $86\% \pm 15\%$ from AGNs with jets oriented toward Earth, called blazars). However, a minor contribution from SFGs cannot be ruled out, as the most active SFGs are known gamma-ray emitters. Above a few hundred GeV, the CGB spectrum cuts off due to gamma-ray interactions with COB and CIB photons, which produce electron-positron pairs. This marks the end of the EBL spectrum.

\subsection{Extragalactic Neutrino and Cosmic-ray Backgrounds}

\begin{figure}[h]
    \centering
    \includegraphics[width=0.49\linewidth]{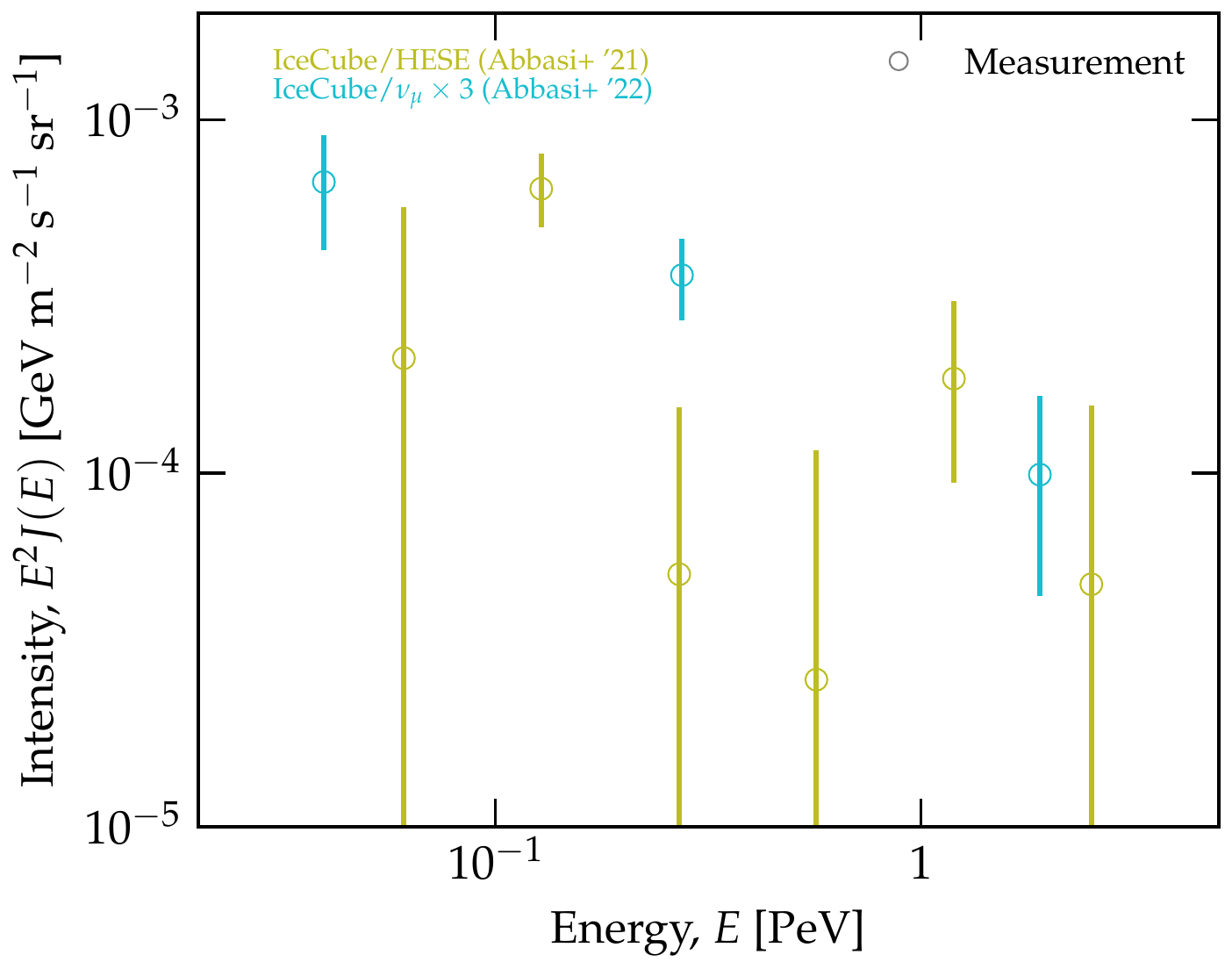}
    \includegraphics[width=0.49\linewidth]{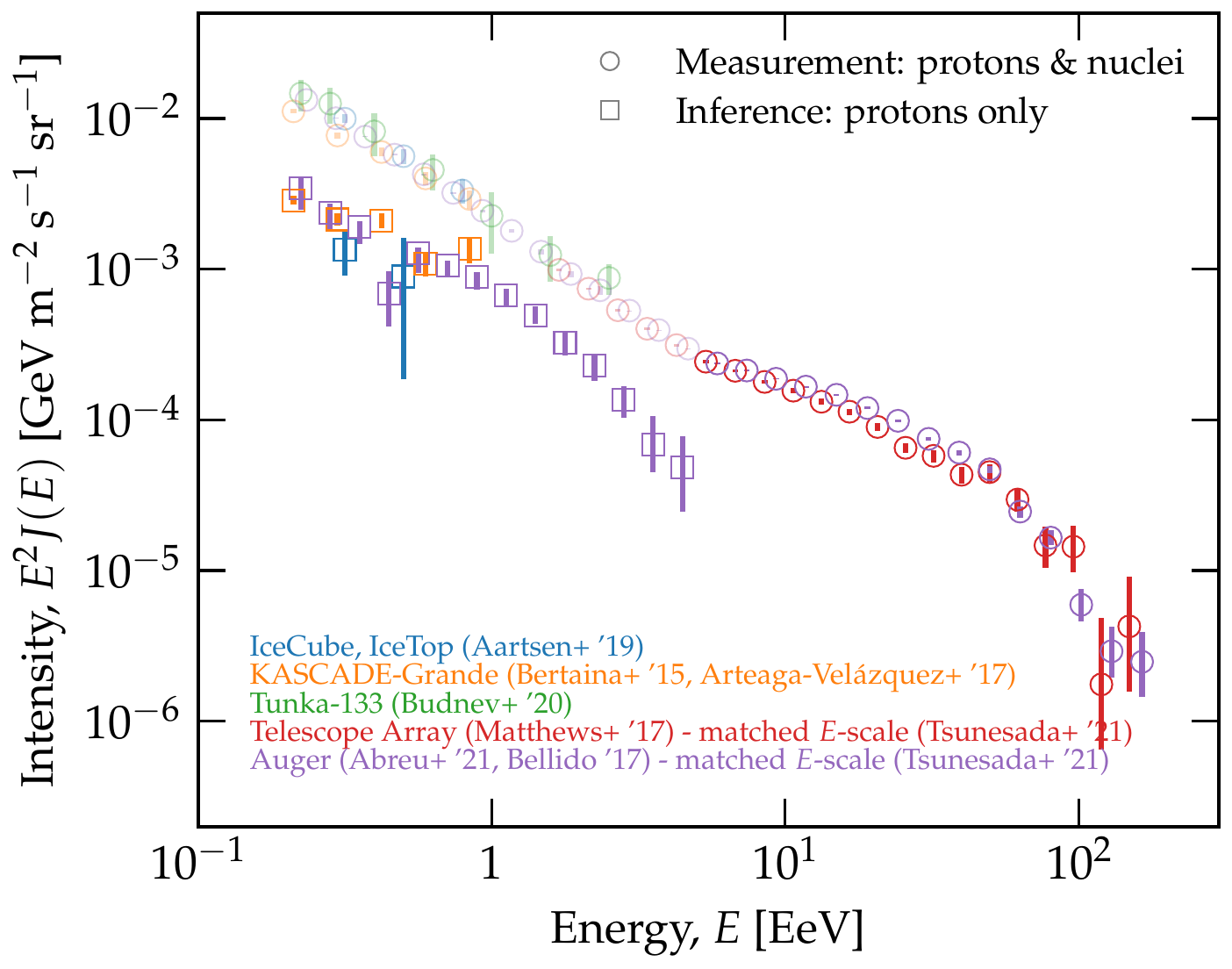}
    \caption{Spectra of the Extragalactic Neutrino (left) and Cosmic-ray (right) Backgrounds.}
    \label{fig:ENCRB}
    \vspace{-0.5cm}
\end{figure}

Beyond TeV energies, the extragalactic spectrum of the universe is dominated by non-photonic messengers. Figure~\ref{fig:ENCRB} shows the spectra of  extragalactic neutrinos ($\nu_e$, $\nu_\mu$, $\nu_\tau$) in the TeV$-$PeV range and atomic nuclei (protons up to ionized oxygen or even iron) in the EeV$-$ZeV range. The ENB exhibits a featureless power-law spectrum, with an energy cutoff constrained to above 500\,TeV \cite{2021PhRvD.104b2002A}. The few indications and the single piece of evidence of point-source excesses are limited to a one-percent contribution to the ENB, so that the majority of the sources remain unresolved. The most robust excess, significant at $4.2\sigma$, comes from the SFG NGC\,1068, which contains a Seyfert-type AGN \cite{2022Sci...378..538I}. The contribution of this AGN sub-population could saturate the ENB \cite{2022PhRvD.106b2005A}, which would link neutrino emission to the cosmic history of accretion.

The ECRB exhibits a more elaborate energy dependence, with spectral slope breaks observed at about 5, 15, and 45\,EeV, known as the 'ankle', 'instep', and 'toe'. As summarised in \cite{2024arXiv241213077B}, these slope changes coincide with an increasing mass of the nuclei that make up the ECRB. Protons dominate around the ankle and nuclei at least as heavy as oxygen are prevalent beyond the toe. As the composition is inferred statistically, the nuclear spectra can hardly be separated. Despite correlations with helium, the proton spectrum is easier to infer and is shown in Figure~\ref{fig:ENCRB} at energies below the ankle \cite{2022ApJ...936...62L}. In this figure, the proton + nucleus spectrum below the ankle is illustrated with a transparency level: the origin of He and CNO nuclei in this region, whether Galactic or extragalactic, remains debated. As with the ENB, identifying the sources of the ECRB remains challenging. While the discovery of a low-amplitude dipole component above the ankle confirms the extragalactic origin of the ECRB, the most robust evidence of anisotropy around the toe, between $4\sigma$ and $4.5\sigma$, suggests an excess in the Centaurus region on an angular scale of ten to twenty degrees, expected from  intervening magnetic fields. This region contains a group of active SFGs and a jetted AGN within the relevant cosmic-ray horizon, a few tens of Mpc. From a purely observational point of view, deciding whether the emission is linked to the ejection of supermassive black holes (e.g., AGNs) or to star formation (e.g., gamma-ray bursts), remains difficult.

\section{Analysis and Discussion: Energy Budgets of Cosmic Backgrounds}

\begin{figure}[b]
    \centering
    \includegraphics[width=0.9\linewidth,page=1]{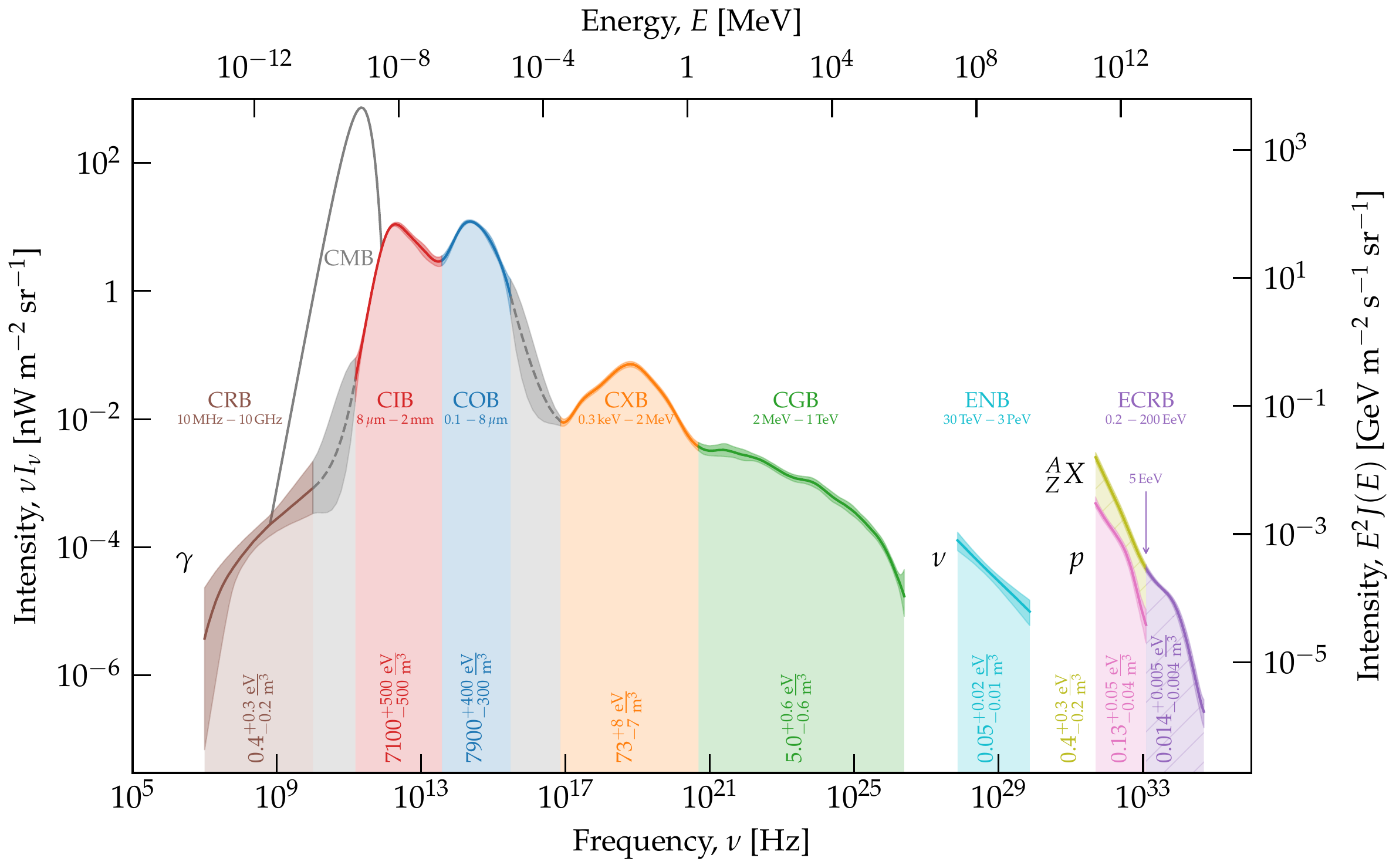}
    \caption{The best-fit spectrum of the multi-messenger extragalactic background.}
    \label{fig:Inventory}
\end{figure}

To determine the radiative balance of extragalactic background components, we model the COB-CIB, CXB-CGB, and ECRB spectra using cubic splines in the $\log\, \nu - \log\, \nu I_\nu$ space, with nodes spaced by half a decade in energy. The ENB and CRB are modeled using power laws. The CRB spectrum follows a power law of $I_\nu \propto \nu^{5/2}$ expected in the synchrotron self-absorption regime, with free frequency break and high-frequency index. The free parameters are determined using a Markov chain Monte Carlo algorithm with normal likelihoods for measurements and half-normal likelihoods for limits. Systematic uncertainties in the CXB-CGB intensity are modeled as lognormal fluctuations of 10\%. Lognormal fluctuations in the energy scale of 15\% are considered for the ENB and ECRB. The best-fit intensity and 68\% confidence contour are presented in Fig.~\ref{fig:Inventory}.

The energy budgets of the COB and CIB shown in the figure (see vertical text) are equivalent to bolometric intensities of $30.1 \pm 1.4$ and $27.0 \pm 1.8\,$nW\,m$^{-2}$\,sr$^{-1}$, respectively. Our data collection and joint analysis confirm with 10\% accuracy equipartition between optical light escaping from galaxies and light absorbed and re-radiated in the infrared. For comparison, assuming a matter-to-light conversion efficiency equal to that of an average star such as the Sun, and using cosmic star-formation models normalized by \cite{2021MNRAS.503.2033K} for a Chabrier initial mass function, we estimate a contribution of $50 \pm 10\,$nW\,m$^{-2}$\,sr$^{-1}$ from stellar nucleosynthesis. An additional 10\% contribution from cosmic accretion could be added to this value. The bolometric intensity of the extragalactic background, measured at approximately 60\,nW\,m$^{-2}$\,sr$^{-1}$, aligns well with the integrated history of astrophysical emission processes, which confirms quantitatively Lord Kelvin's initial inference and Harrison's theoretical confirmation. An equally satisfying back-of-the-envelope calculation can be made for the CGB, by considering the ratio of the kinetic luminosity function of AGN jets to cosmic accretion as well as the kinetic-to-radiation conversion efficiency within astrophysical jets.

Another point of comparison stems from the ENB energy budget, which is $50^{+20}_{-10}\,$meV\,m$^{-3}$ between 30\,TeV and 3\,PeV. These neutrinos originate from the interaction of protons with energies approximately twenty times higher. Assuming that half of these protons interact, and a neutral-to-charged-pion ratio of one (see \cite{2018PrPNP.102...73A}), the proton energy density would lie between 200 and 400\,meV\,m$^{-3}$ within the $0.6-60\,$PeV proton energy range, i.e., below the ECRB observation range. This order of magnitude is comparable to the proton energy density observed at $130^{+50}_{-40}\,$meV\,m$^{-3}$ between 200\,PeV and 5\,EeV, with uncertainties primarily driven by the energy scale. In this context, the neutrino candidate detected in the Mediterranean Sea with an energy ranging from 0.1 to 0.8 PeV \cite{2025Natur.638..376K} is of particular interest, since the proton that produced it should be located near the ankle. The neutral pions generated by such protons decay into gamma rays, which cascade to lower energies due to interactions along the line of sight. Using the analytical cascade model of \cite{2016PhRvD..94b3007B}, we estimate the energy density of the gamma-ray counterpart of the ENB to be $25-50\,$meV\,m$^{-3}$ between 10\,GeV and 1\,TeV. This is one order of magnitude below the CGB observed in this range ($300 \pm 40$\,meV\,m$^{-3}$). These comparisons of observed and expected intensities suggest links between multi-messenger extragalactic backgrounds. However, given the accessible energy ranges and current measurement uncertainties, observationally constraining these links may only become feasible for the next generation of enthusiasts of the dark-night-sky paradox. Extending the population synthesis models developed for EBL to other extragalactic messengers could be another avenue for constraining the power engines behind the most energetic particles in the Universe.\vspace{-0.2cm}

\setlength{\bibsep}{1pt}
\linespread{1}\selectfont

%\bibliography{biblio}

\providecommand{\href}[2]{#2}\begingroup\raggedright\endgroup

\end{document}